\documentstyle[aps,preprint]{revtex}
\newcommand{\be}{\begin{equation}}
\newcommand{\ee}{\end{equation}}
\newcommand{\bea}{\begin{eqnarray}}
\newcommand{\eea}{\end{eqnarray}}

\begin{document}
\bibliographystyle{prsty}

\draft 


\title{Parametric Forcing of Waves with Non-Monotonic Dispersion Relation: 
Domain Structures in Ferrofluids?}

\author{David Raitt\\
SCRI, Florida State University B-186\\
400 Dirac Science Center Library, Tallahassee, FL 32306\\
Hermann Riecke \\
Department of Engineering Sciences and Applied Mathematics\\
Northwestern University, Evanston, IL 60208\\
}
\maketitle 

\begin{abstract}

Surface waves on ferrofluids exposed to a dc-magnetic field exhibit
a non-monotonic dispersion relation. The effect of a parametric driving on such
waves is studied within suitable coupled Ginzburg-Landau equations. 
Due to the non-monotonicity the neutral curve for the excitation of standing waves
can have up to three minima.  
The stability of the waves with respect to long-wave perturbations is determined
$via$ a phase-diffusion equation. It
shows that the band of stable wave numbers can split up into two or three
sub-bands. The resulting competition between the 
 wave numbers corresponding to the respective sub-bands
leads quite naturally to patterns consisting of multiple domains of standing waves
which differ in their wave number. The coarsening dynamics of such domain structures is addressed.
{\centerline \today}

\end{abstract}
\pacs{05.45.+b, 05.90.+m, 47.20. Tg}

\section{Introduction.} 
\label{s:intro}
Spatial patterns have been studied extensively over the past years, 
the classic 
examples being Rayleigh-B\'enard convection, Taylor vortex flow, as well as structures
arising in directional solidification \cite{CrHo93}. It has been well established that
long-wavelength perturbations of such steady patterns exhibit diffusive phase
dynamics \cite{PoMa79} with the phase-diffusion coefficient changing sign at the Eckhaus instability \cite{Ec65,KrZi85}.
Thus, after the decay of transients one-dimensional steady patterns usually
relax to a strictly periodic pattern. 

More recently, it has been pointed out, however, that this
need not be the case in general; there are situations in which the final state consists
of a number of domains with different wave numbers \cite{BrDe89,Ri90a,DeLe90,BrDe90,Ri90}. This can occur if the phase-diffusion coefficient becomes
negative such that the band of stable wave numbers is split into two parts. Within
each domain the wave number is then in one of the two stable sub-bands. Experimentally,
such domain structures have been observed in Rayleigh-B\'enard convection
in a narrow channel \cite{HeVi92}. 
So far it has, however, not been firmly established whether the origin of these states is
in fact due to a splitting of the stable band since the phase-diffusion coefficient has
not been measured in this regime. From a theoretical point of view this experiment
is difficult to analyze since the domain structures arise only at very large
Rayleigh numbers which require full numerical simulations of the three-dimensional
Navier-Stokes equations. 
 
Domain structures can be viewed as arising from the competition between different
wave numbers. For their investigation it is therefore natural to turn to pattern-forming systems which have
a neutral curve with two or more minima corresponding to the competing wave numbers.
Domain structures are then expected to arise close to threshold where
they may be described analytically within a small-amplitude theory 
\cite{Ri90a,Ri90,RaRi95,RaRi95a}. In addition, in 
this regime interesting locking \cite{CoEl87} and 
coarsening phenomena are possible which are not
expected at large amplitudes \cite{RaRi95,RaRi95a,PeKr86,BoZi88,BoKa90}.

So far, not many physical systems have been identified which exhibit a neutral curve
with two minima. Note that within the framework of Ginzburg-Landau equations
it is not sufficient
if two different modes with different wave numbers go unstable at the same 
value of the control parameter. Since the stability of the domain structures 
relies either on phase conservation
\cite{BrDe90,Ri90a,RaRi95a} or on a locking-in of interacting fronts \cite{RaRi95a}
(see also below) both minima have to correspond to the same mode. 

For traveling waves a neutral curve with a double minimum has recently been identified 
in convection of a conducting fluid in a rotating annulus in the presence of a 
magnetic field \cite{BuFi93}. This system is, however, not easily accessible in 
experiments.

In this paper we show that neutral curves with multiple minima, and the resulting
domain structures, may be obtained quite naturally by a 
parametric driving of waves with {\it non-monotonic} dispersion relation. 
Then there exists a range of frequencies
in which modes with different wave numbers resonate simultaneously
with the driving. The dispersion relation becomes non-monotonic when the group velocity
changes sign. This is the case in
spiral vortex flow between counter-rotating cylinders in certain parameter regimes
 \cite{TaEd90}. However, the parametric forcing of 
spiral vortex flow has turned out to be non-trivial due to the appearance of Stokes
layers \cite{TeAn96}. If the parametric driving can be applied $via$ a bulk forcing
(e.g. using electric or magnetic fields) its effect is considerably stronger. 
This suggests considering surface waves on ferro-fluids. They
 exhibit a non-monotonic dispersion relation if they are exposed to a sufficiently
strong dc-magnetic field \cite{MaGr96}. 
In fact, the non-monotonicity is a precursor of the Rosensweig
instability \cite{Ro85a,Ro87}. Quite analogously, surface waves on conducting 
and dielectric fluids in the presence of an electric field exhibit a non-monotonic
dispersion relation leading to 
an instability for larger fields \cite{TaMc65,HeMi90,SuCh93,El94}. 

The organization of the paper is as follows.
To describe parametrically driven waves for small amplitudes we 
introduce in sec.\ref{s:gl} suitable coupled complex Ginzburg-Landau
equations. In sec.\ref{s:diff} we study analytically the long-wave stability of 
standing waves which are parametrically 
excited by a periodic forcing (e.g. by an ac-magnetic or an 
ac-electric field). 
In the expected parameter regime the band of stable wave numbers separates into 3 sub-bands. 
The numerical
simulations presented in sec.\ref{s:num} confirm the existence and stability of structures consisting of domains
with different wave number. In addition, we present numerical results for the coarsening of 
arrays of domains within a single, $4^{th}$-order Ginzburg-Landau equation
which can be derived from the coupled Ginzburg-Landau equations under suitable conditions.
 
\section{The Model} 
\label{s:gl}
For small amplitudes and
small damping parametrically driven waves can be described by coupled 
Ginzburg-Landau equations \cite{RiCr88,Wa88,RiSi94}. 
A crucial ingredient for determining the linear part of these equations
is the dispersion relation of the waves. For surface waves on an inviscid
ferro-fluid of infinite depth in the presence of a dc-magnetic field it is given by \cite{Ro85a}
\be
\omega^2(q)=gq+\frac{\sigma}{\rho} q^3 -\frac{1}{\rho(1/\mu_0+1/\mu)}M_0^2 q^2.
\label{e:disp}
\ee
Here $g$ is the gravitational acceleration, $\sigma$ the surface tension, $\rho$ the
density of the fluid, $\mu$ the permeability and $M_0$ the magnetization of the fluid.
When $M_0^2$ is increased beyond $M_c^2 \equiv \sqrt{3g\sigma \rho}(1/\mu_0+1/\mu)$ the 
dispersion relation becomes non-monotonic and is given by a cubic polynomial in the
vicinity of the inflection point. 
To capture this cubic dispersion relation third spatial derivatives are retained
in the Ginzburg-Landau equations,
\bea
\partial_t A+v \partial_x A&=&d \partial^2_x A + f \partial^3_x A+a A+b B+
c A |A|^2 + (n-c)|B|^2A, \label{e:cglA}\\
\partial_t B-v \partial_x B&=&d^* \partial^2_x B + f^* \partial^3_x B+a^* B+b A+
c^* B |B|^2 +(n^*-c^*)|B|^2B. \label{e:cglB}
\eea

Physical quantities like the surface height $h$ are described in terms of the complex amplitudes
as 
\be
h=\delta e^{iq_0\hat{x}} \left( A(x,t) e^{-\omega_e/2 \hat{t}} + 
B(x,t) e^{\omega_e/2 \hat{t}}\right) + c.c. + h.o.t., \qquad \delta \ll 1,
\ee
where the amplitudes $A$ and $B$ depend on slow space and time coordinates, 
$t=\delta^2 \hat{t}$ and $x=\delta \hat{x}$. 
 The parametric driving, which can be achieved with an additional ac-magnetic field, enters
the equations $via$ the linear coupling terms $bA$ and $bB$, respectively. 
Its strength is proportional to the coefficient $b$ \cite{RiCr88,Wa88},
which can be chosen real. All other coefficients are in general complex. 
 The second control parameter is the detuning between the 
frequency $\omega_e$ of the external driving and the natural frequency $\omega_0$
of waves with wave number $q_0$,
$\omega_0=\omega_e/2+a_i-\alpha a_r$ with $\alpha$ being an $O(1)$-quantity and 
$a\equiv a_r+ia_i$. The carrier wave number $q_0$ is chosen to 
be that wave number for which the dispersion
relation (\ref{e:disp}) has an inflection point with zero slope.
This occurs for $M_0=M_c$. At this point
the linear group velocity $v_r$ as well as the quadratic 
dispersion term $d_i$
vanish. As long as $M_0$ is close to $M_c$ these dispersive terms
are therefore small and it is consistent to keep also the third-order term which gives
the cubic dispersion relation. In all of the following we assume that 
the viscosity of the fluid
is low. The dissipative terms are then small with the leading order term being $a_r$, 
allowing us to neglect the imaginary part of $f$. We keep, however, $v_i$ and $d_r$,
 which give the linear and the quadratic dependence of the damping on the wave number, 
although they are also of higher order than $a_r$. In this communication we do not attempt
to make quantitative predictions for a specific experimental system. Therefore we
do not calculate the coefficients of (\ref{e:cglA},\ref{e:cglB}). Instead we present 
results that should be typical for the parametric driving of waves with non-monotonic
dispersion relation. 

The neutral-stability curve, at which  the basic state (with flat surface) 
becomes unstable,  is given by 
\be
b^2=\left|R\right|^2 \qquad \mbox{ with } \qquad R=a-ivq-dq^2-ifq^3.\label{e:nsc1}
\ee
The resulting neutral-stability curve is sixth-order in $q$, and
can therefore have up to three minima. Depending on the parameters, the
absolute minimum of the neutral-stability curve can be any one
of the three. Of course, the situation with a single minimum can be recovered as well.
Examples of neutral curves with multiple minima are given by the solid lines in 
figs.\ref{f:nseckcurves}(a)-(c). There the effect of changing the driving frequency, 
i.e. the detuning $a_i$, is demonstrated. The difference between half the external
frequency and the frequency of a wave with wave number $q$ is indicated by the dotted line.
The damping is also taken to be wave-number dependent ($v_i$ and $d_r$ non-zero). As expected the
neutral curve exhibits local minima at the resonance wave numbers. As the detuning is
changed from negative to positive values the 
absolute minimum shifts from the resonance at high wave number to that at low wave number. 
 
\section{Linear Stability and Phase Diffusion} 
\label{s:diff}
The multiple wells in the neutral curve suggest that the band of stable wave numbers  also
splits into separate sub-bands. This is a prerequisite for stable domain structures
to exist. We therefore determine the linear stability of the nonlinear waves with respect to 
long-wave perturbations. This is done most efficiently by deriving the phase-diffusion equation
\cite{PoMa79,Ri90a},
 \be
\partial_T\phi(X,T)={\cal D}(q)\,\partial_X^2\phi(X,T).
\label{e:pheq1}
\ee
In this description the amplitudes $A$ and $B$ are proportional to 
$S(q)\,e^{i\phi/\epsilon}+ O(\epsilon)$
 where the amplitude $S$ of the wave satisfies
\be
b^2=\left|n\right|^2S^4+2n^*RS^2+\left|R\right|^2
\label{e:s2}
\ee
and $X=\epsilon x$ and $T=\epsilon^2 t$ are superslow scales. The phase $\phi$ gives the
local wave number via $q=\partial_X\phi$.
In order to simplify the expression for ${\cal D}(q)$ the following notation is
introduced
\be
a*b=a_rb_r+a_ib_i, \qquad a\bullet b=a_rb_i+a_ib_r.
\label{e:notation1}
\ee
The phase diffusion coefficient ${\cal D}(q)$ can then be written as $D(q)/\tau(q)$ with
\bea
D(q)&=&-\left\{4\left(3c_ifq+c*d\right)\left|n\right|^2\right\}S^6\nonumber\\
&-&\left\{-18c*n^*f^2q^4+24\left(Im\left(d^*cn\right)\right)fq^3\right.
\nonumber\\
& &+\left[-12\left(Re\left(v^*cn\right)\right)f
+8\left(d*d^*c*n^*+2d_rd_ic\bullet n\right)\right]q^2\nonumber\\
& &+\left[6\left(R_i\left|n\right|^2+2c_in*R\right)f
+8\left(Im\left(cd^*nv^*\right)\right)\right]q\nonumber\\
& &\left.+2\left[d*R\left|n\right|^2+2n*Rc*d-v*v^*c*n^*-
2v_rv_ic\bullet n\right]\right\}S^4\nonumber\\
&-&\left\{9\left[-R*n^*+2R_ic_i\right]f^2q^4
+12\left[c_id_rR_r+\left(c_id_i+c*d\right)R_i+Im\left(Rnd^*\right)
\right]fq^3\right.\nonumber\\
& &+\left[6\left(-c_iv_iR_r+\left(c_iv_r+c\bullet v^* \right)R_i
-Re\left(Rnv^*\right)\right)f\right.\nonumber\\
& &\left.\hspace*{.25in}+8c*dd*R+4d*d^*n*R^*+
8d_rd_in\bullet R\right]q^2\nonumber\\
& &+\left[4\left(v^*\bullet Rd*n+v*Rd^*\bullet n+c*dv^*\bullet R+d*Rv^*\bullet c\right)+6R_ifn*R\right]q\nonumber\\
& &+\left[2\left(d_rn_rR_r^2+d_in_iR_i^2+d\bullet nR_rR_i\right)\right.\nonumber\\
& &\hspace*{.25in}\left.\left.-2\left(c+n\right)\bullet Rv_rv_i
+\left(2c_iR_i-n*R^*\right)v_r^2 +\left(2c_rR_r+n*R^*\right)v_i^2\right]
\right\}S^2\nonumber\\
&-&\left\{9R_i^2f^2q^4 +12d*RR_ifq^3+\left[6R\bullet v^*R_if+
4\left(d*R\right)^2\right]q^2+4d*RR\bullet v^*q+
\left(R\bullet v^*\right)^2\right\} \label{e:D}
\eea
and
\be
\tau(q)=-4c_r\left |n\right|^2S^6-\left(4\,c_rn*R+2R_r\left|n\right|^2\right)S^4
-2R_rn*RS^2.
\label{e:tau}
\ee
Here $Re(a)$ and $Im(a)$ denote the real and imaginary part of $a$, respectively.


The stability limit (Eckhaus boundary) of the spatially periodic waves 
is given by ${\cal D}(q)=0$ and is denoted by the dashed lines in 
figs.\ref{f:nseckcurves}(a)-(c).
From (\ref{e:nsc1}) and (\ref{e:D},\ref{e:tau}) one can see that 
if the imaginary parts of the
various parameters are zero then the neutral-stability and the Eckhaus curves are 
symmetric around $q=0$. Even if the neutral curve has multiple minima and consequently
the band of stable wave numbers is separated into sub-bands, these sub-bands merge
for large forcing. This can be seen directly from (\ref{e:D},\ref{e:tau}) by considering
the limit $S \rightarrow \infty$. For $c_i=0$ the leading-order term 
$\left\{-4\left(3c_ifq+c*d\right)\left|n\right|^2\right\}S^6$ 
of $D(q)$ is positive since $c_r <0$ and
$d_r>0$. Thus, ${\cal D}$ becomes positive for large $S$ independent of the other
coefficients, since $\tau >0$ for large $S$.

The same consideration shows that for general coefficients, i.e. for $c_i \ne 0$ and 
$f \ne 0$, the diffusion coefficient ${\cal D}$ always becomes negative for large $S$ for 
some range of $q$. For the parameters chosen in figs.\ref{f:nseckcurves}(a)-(c) this 
occurs for $q>0$. Thus, for $a_i=-0.7$ one obtains two stable sub-bands which do not
merge for large forcing. With increasing $a_i$ two additional resonance arise for smaller
$q$ leading to a quite flat neutral curve. Strikingly, the left stable sub-band splits up into
two bands leading to three separate bands right at onset (fig.\ref{f:nseckcurves}(b)). 
Upon a further increase to $a_i=0.5$ the large-$q$ sub-band merges with the central one
for small forcing. For large forcing they remain separated. The merging of 
the central band with the left band occurs now only at larger values of the forcing
$b$ (fig.\ref{f:nseckcurves}(c)).

\section{Numerical Simulations} 
\label{s:num}
\begin{figure}
\begin{picture}(420,200)(0,0)
\put(-20,-45) {\includegraphics{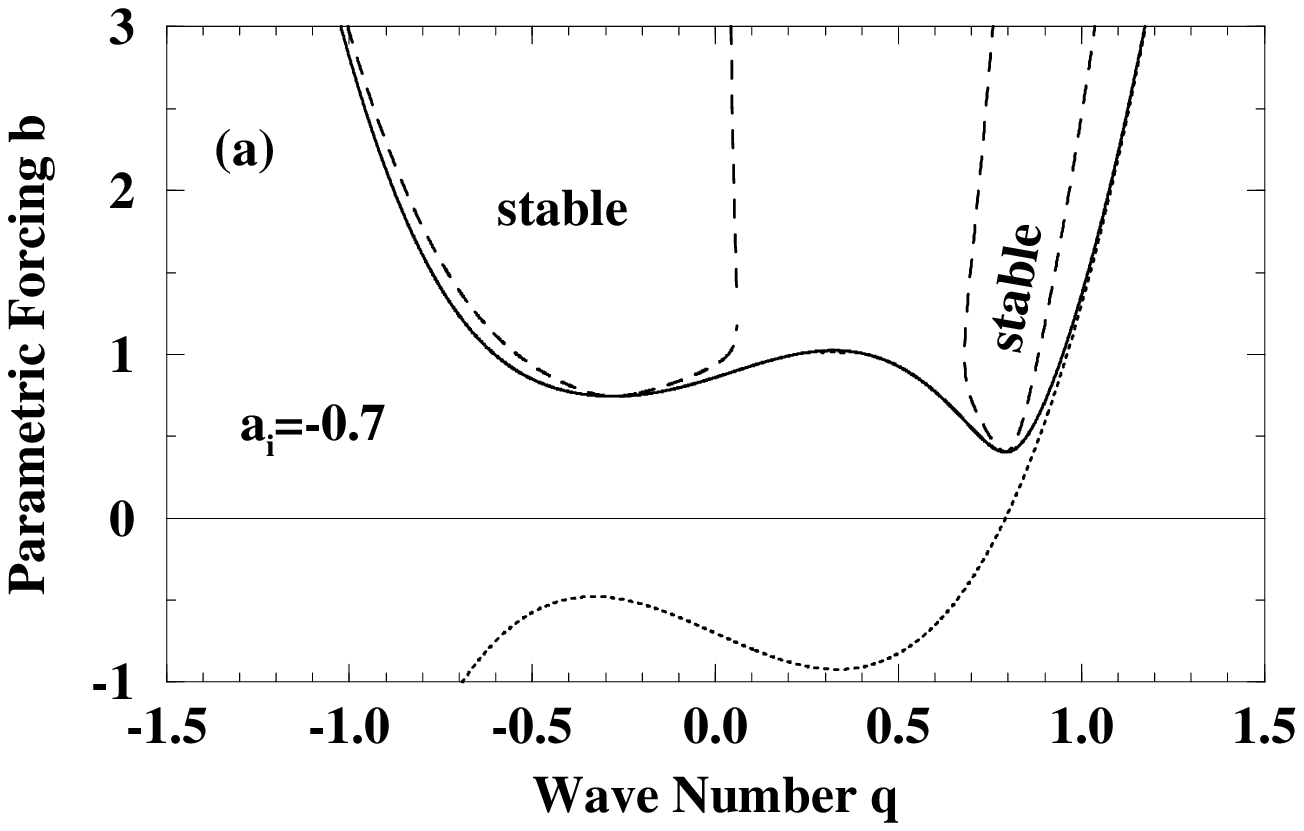}}
\end{picture}
\begin{picture}(420,240)(0,0)
\put(0,-45) {\includegraphics{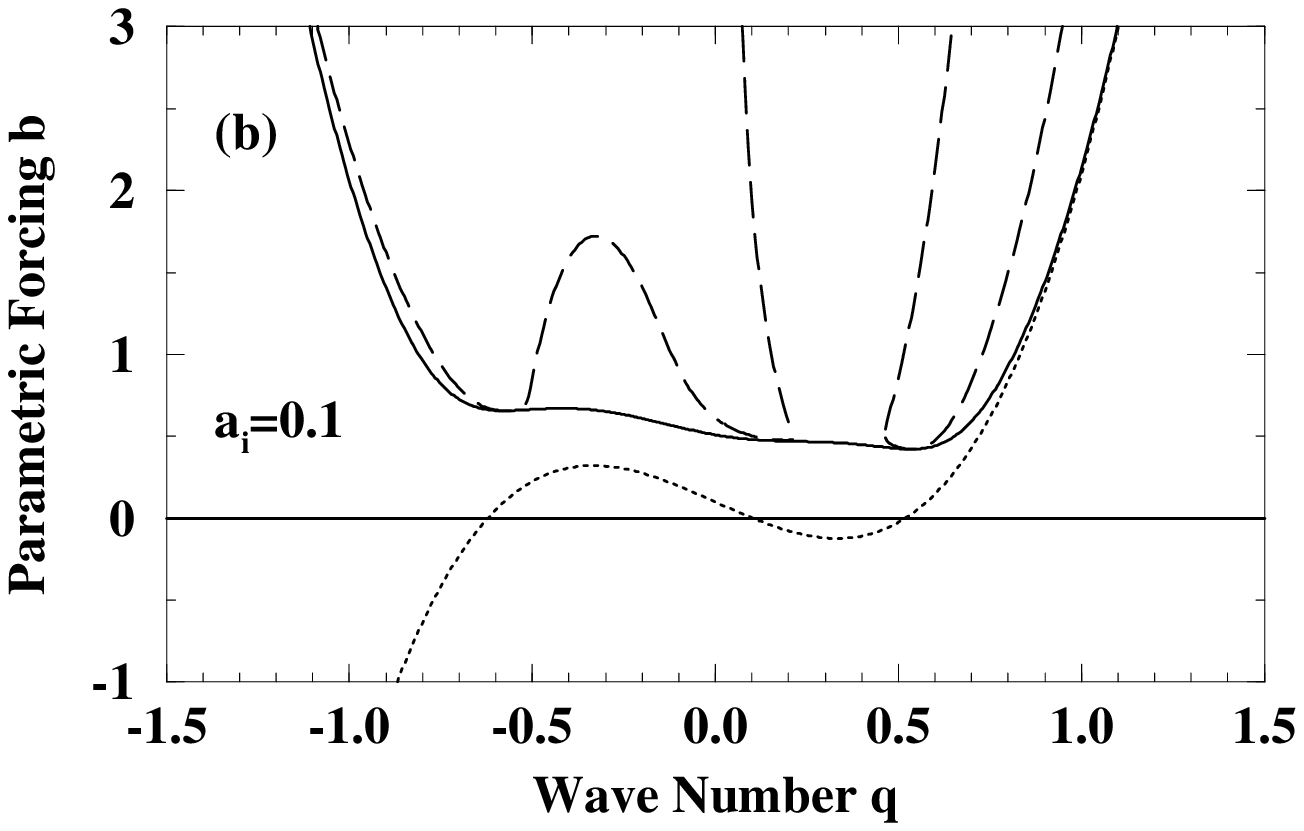}}
\end{picture}
\begin{picture}(420,240)(0,0)
\put(0,-45) {\includegraphics{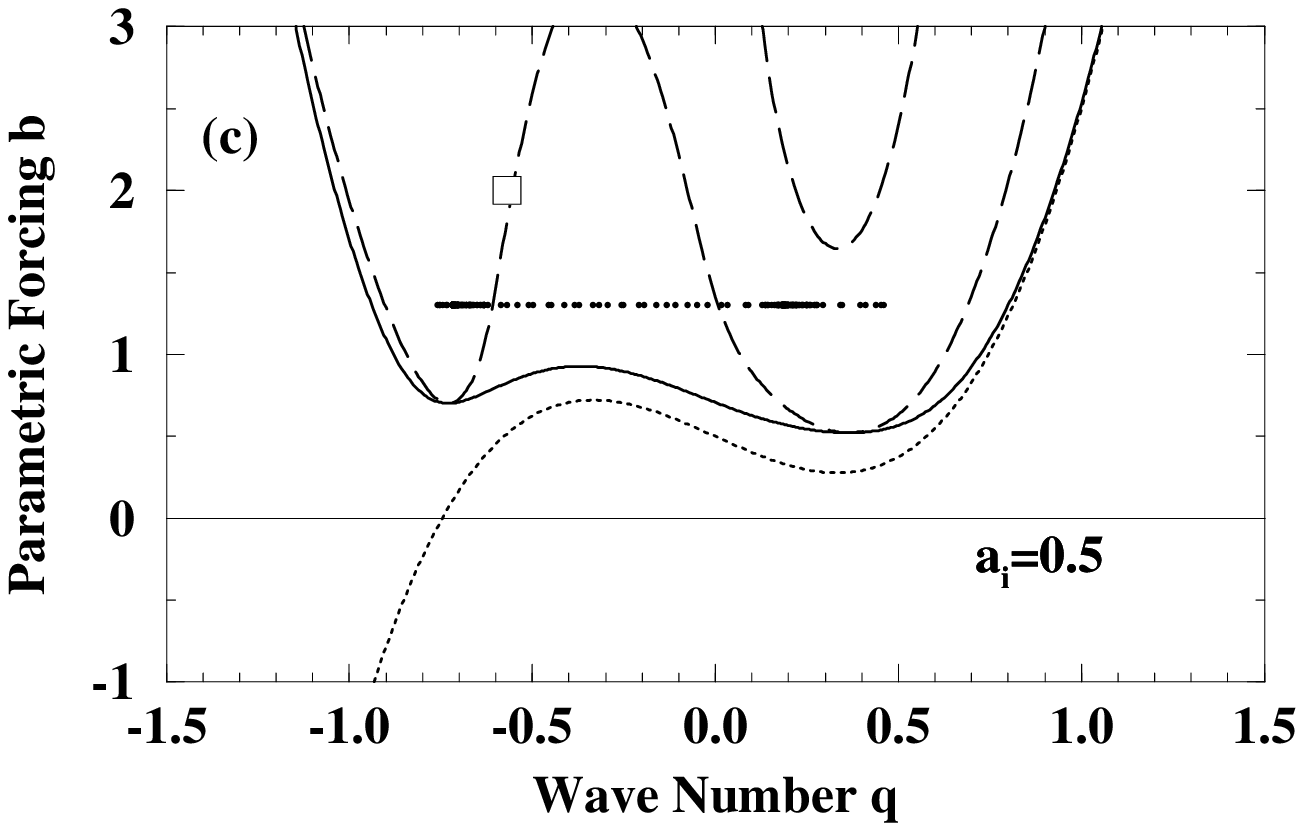}}
\end{picture}
\caption{The Eckhaus (dashed line) and neutral-stability (solid line) curves for 
eqs.(\protect{\ref{e:cglA}},\protect{\ref{e:cglB}})
 for $a_r=-0.5$, $d=0.1$, $v=1+0.2i$, $c=-1-0.5i$, $n=-3+2i$ and
various values of the detuning $a_i$ as indicated. The dotted line in (c) 
indicates the wavenumber distribution of the structure shown in
fig.\protect{\ref{f:typsol}}. 
\protect{\label{f:nseckcurves}}
}
\end{figure}

\begin{figure}
\begin{picture}(420,240)(0,0)
\put(0,-45) {\includegraphics{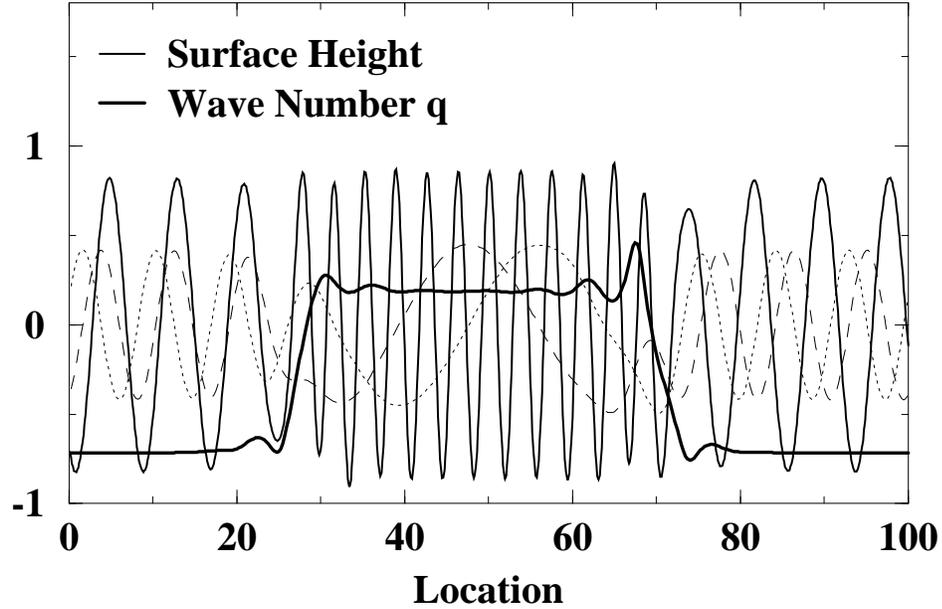}}
\end{picture}
\caption{A typical domain structure arising from 
eqs.(\protect{\ref{e:cglA}},\protect{\ref{e:cglB}})
 for $a_r=-0.5+0.5i$, $d=0.1$, $v=1+0.2i$, $c=-1-0.5i$, $n=-3+2i$. The thin solid line
gives $Re(e^{iq_cx}(A+B))$ with $q_c=1.5$. The short and long dashed lines indicate the
real and imaginary part of $A$, respectively.
 \protect{\label{f:typsol}}
}
\end{figure}

To show that the complex stability regions shown in 
figs.\ref{f:nseckcurves}(a)-(c) indeed lead to stable domain structures consisting of 
domains with large and small wave numbers we have solved (\ref{e:cglA},\ref{e:cglB})
numerically using a finite-difference code with Crank-Nicholson time-stepping. 
A typical domain structure is shown in fig.\ref{f:typsol}.
It was obtained for parameter values $a=-0.5+0.5i$, $b=1.3$, $c=-1-0.5i$, $d=0.1$, 
$f=-3$, $n=-3+2i$, and $v=1+0.2i$.
 These are the parameter values that correspond to fig.\ref{f:nseckcurves}(c)
and the location of this domain structure in $(q,b)$-space is indicated by the
dotted line on that figure.
The domain structure itself consists of a region with
$q\approx 0.19$ in the center of the figure and $q\approx -0.72$ at the boundaries
(thick line). These
wave numbers correspond to the two Eckhaus-stable wells for the selected parameter
values. Between the two domains there is a sharp transition. This domain structure is 
numerically stable. The thin line gives $Re(exp(iq_cx)\{A + B\})$ with $q_c=1.5$
and is intended to give an impression of 
a typical surface deformation in this regime. The broken lines give the real and 
imaginary part of $A$. The dots in fig.\ref{f:nseckcurves}(c) 
give the wave-number
distribution of the solution shown in fig.\ref{f:typsol}. The higher density of dots
near $q=-0.7$ and $q=0.2$ shows that over most of the system the local wave number 
is within one of the two stable regimes.

From an experimental point of view an important question is how one obtains these
domain structures. Clearly, an adiabatic increase of the periodic forcing in the
frequency regime in which the neutral curve has two different minima will not be
successful since the emerging pattern will always be periodic with the wave number corresponding to the deeper minimum. Instead, one has to increase the forcing suddenly from
values below threshold to a value for which
the wave-number band consists of at least two sub-bands. Even then one may still predominantly get
patterns with the wave number corresponding to the deeper minimum. Alternatively, 
one can change the forcing frequency at fixed supercritical forcing amplitude in the regime
in which the stable band is split. The change in frequency shifts the band of stable wave
numbers and eventually the initially stable wave becomes unstable. If its wave number
hits the part of 
the stability boundary which faces the other sub-band (as marked by the open square in 
fig.\ref{f:nseckcurves}(c)) it is expected that the
instability will not lead to a phase slip but to domain structures.

\begin{figure}
\begin{picture}(420,520)(0,0)
\put(-80,-100) {\includegraphics{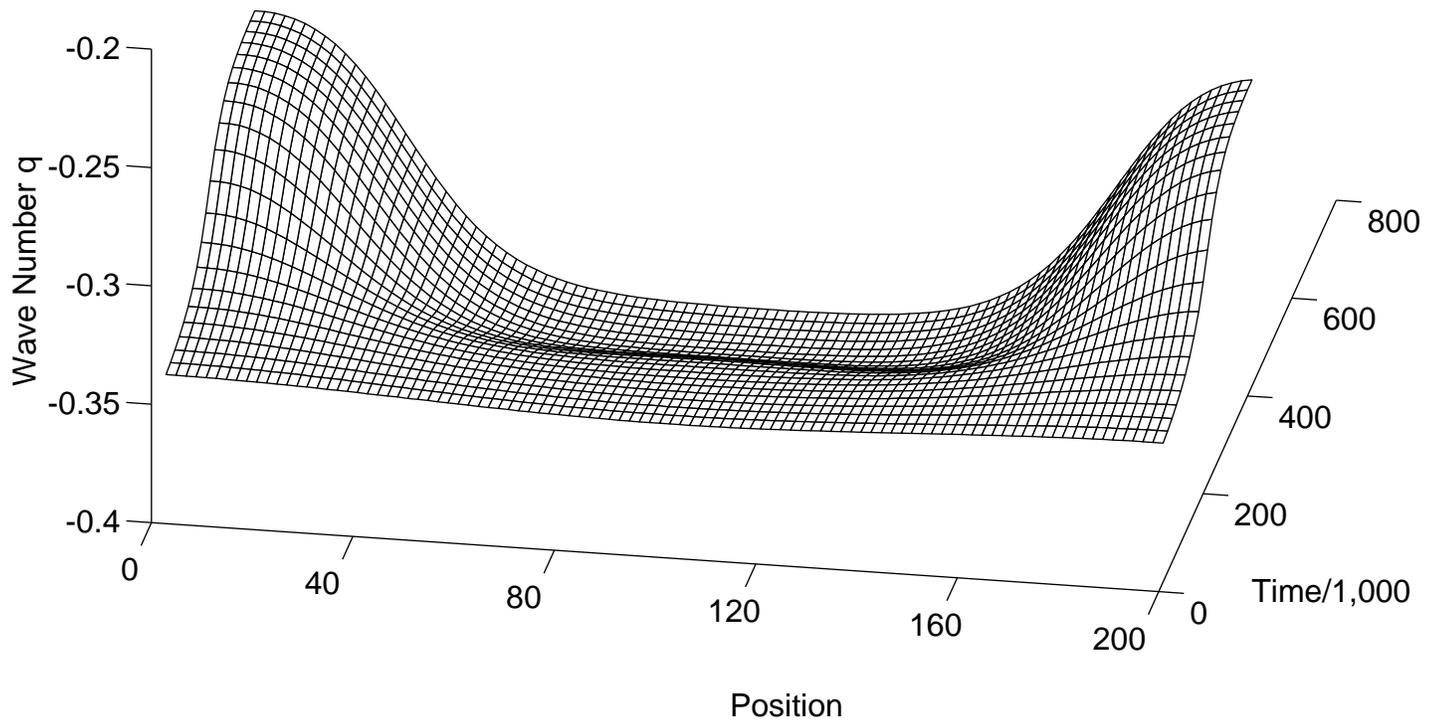}}
\end{picture}
\caption{Temporal evolution of the local wave number. The initial wave number
 is $q\approx-0.34$. The forcing is $b=1.7$. The remaining parameters
are $a=-0.5+0.1i$, $d=0.1$, $v=1+0.2i$, $c=-1-0.5i$, $n=-3+2i$,
 $f=-3$. 
}
\protect{\label{f:exper1}}
\end{figure}

The best approach is presumably to
 prepare a periodic pattern in the single-well regime, i.e. for $M < M_c$, with
the wave number (i.e. the frequency) chosen such that it falls
into the unstable region between two sub-bands
once the dc-magnetic field is suddenly increased to reach $M > M_c$. The numerical result of 
such a protocol is shown in fig.\ref{f:exper1}. It gives the space-time diagram for the
local wave number and clearly shows how the pattern separates into domains with 
different wave numbers after a jump from $v_r >0$ to $v_r <0$ which renders the initial
wave number in the region of instability between the left two sub-bands of
 fig.\ref{f:nseckcurves}(b) ($b=1.7$).
In the simulation the initial condition was perturbed with a long-wave modulation of the
wave number in order to trigger the long-wave Eckhaus instability. Note that the 
initial condition was placed very close to the top of the hump of the Eckhaus curve in 
order to insure that the fastest growing mode has a long wavelength leading to a 
single low-wavenumber domain. Thus, the diffusion coefficient is only weakly negative and
the evolution is extremely slow.

\begin{figure}
\begin{picture}(420,520)(0,0)
\put(-80,-100) {\includegraphics{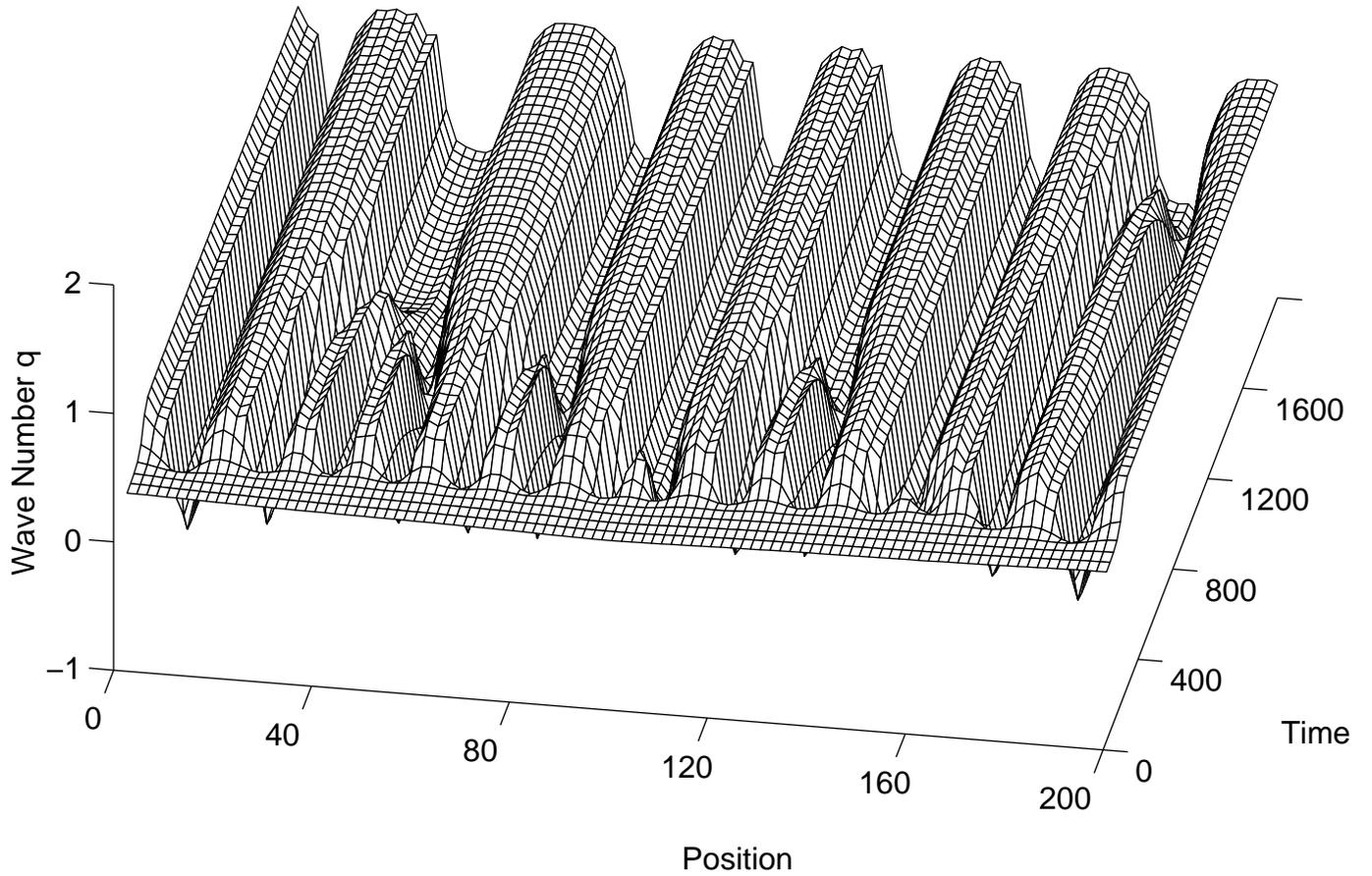}}
\end{picture}
\caption{Temporal evolution of the local wave number. The initial wave number
 is $q=0.28$ and $b=2$. The remaining parameters are as in 
fig.\protect{\ref{f:exper1}}. The fastest growing mode has a short wavelength and
leads to a large number of small domains which then undergo a coarsening process.
}
\protect{\label{f:exper2}}
\end{figure}

A second simulation is shown in fig.\ref{f:exper2}. 
Here the initial wave number is $q=0.28$ 
and $b=2$, i.e. the wave number falls between the two sub-bands which are completely separated
from each other. The evolution of the wave number is strikingly different from
fig.\ref{f:exper1}. Although the initial perturbation was the same as in fig.\ref{f:exper1}
large oscillations in the wave number arise after a short time. Apparently, in this regime
the fastest growing mode has a short wavelength. Presumably, 
this represents the usual shift to shorter wavelengths when the Eckhaus boundary is
exceeded substantially. In principle it could, however, also indicate
 an additional short-wavelength instability not captured in the 
phase equation (\ref{e:pheq1}). 

Fig.\ref{f:exper2} shows that the side-band instability can lead - at least initially -
to quite complex arrays of domains. As can be seen in that figure, the domains evolve
slowly over time in a coarsening process in which adjacent domain walls (fronts)
annihilate each other thereby reducing the number of domains. Thus,   
the stability of a complex structure like that shown in fig.\ref{f:exper2} depends crucially
on the interaction between the domain walls. Oscillations in the wave number
-- as apparent in fig.\ref{f:typsol} -- 
suggest that the interaction could be oscillatory in space, thus allowing for
a discrete set of equilibrium distances between the domain walls \cite{CoEl87}. 
This question has
been addressed previously within the framework of a single Ginzburg-Landau equation
with fourth-order spatial derivatives, which models a neutral curve with two equal minima
 \cite{PeKr86,BoKa90,RaRi95,RaRi95a}. As discussed there, no
locking is possible within the phase equation (\ref{e:pheq1}). Locking
 can therefore arise only in regimes in which the phase 
equation breaks down, as is, for instance, the case very close to threshold.

Close to threshold the coupled Ginzburg-Landau equations 
(\ref{e:cglA},\ref{e:cglB}) can be reduced to a single, real Ginzburg-Landau equation. 
To reduce the numerical effort we therefore consider now the equation
\be
\frac{\partial{\cal A}}{\partial \tilde{T}}=D\frac{\partial^2{\cal A}}{\partial\tilde{X}^2}+
iD_3\frac{\partial^3 {\cal A}}{\partial \tilde{X}^3}-
\frac{\partial^4 {\cal A}}{\partial \tilde{X}^4}+\Sigma{\cal A}-\left|{\cal A}\right|^2{\cal A}.
\label{e:gl4}
\ee
If the neutral curve (\ref{e:nsc1})
is well approximated by a quartic polynomial and if its wells
are not too deep (\ref{e:gl4})   
is sufficient to capture the multiple minima \cite{Ri90a}.
In (\ref{e:gl4}) the complex amplitude ${\cal A}$ is proportional 
to the amplitudes $A$ and $B$ of
the left- and right-traveling wave components of the standing wave $|A|=|B|$
 and the cofficients are all real.
The slowness of the new slow scales $\tilde{X}$ and $\tilde{T}$ is
 related to the distance from threshold. We expect that the results obtained from 
(\ref{e:gl4}) 
carry over to the full equations (\ref{e:cglA},\ref{e:cglB}). 
In addition, we restrict ourselves to the simpler,
symmetric case $D_3=0$. This case has also been studied in the context of two-dimensional
zig-zag patterns. There the locking discussed below has been identified previously
\cite{PeKr86,BoZi88,BoKa90}.

In the general, asymmetric case $D_3 \ne 0$ the minima of the neutral curve have different
depth. Continuity suggests that the small-amplitude waves near the higher minimum experience
then a short-wavelength instability which takes them directly to the fastest growing
mode near the center of the other stability band. For large values of $\Sigma$ the stable
band connected with the higher minimum is known to close \cite{Pr91}. We therefore
expect the results of the symmetric case to carry over to the general case only
for intermediate values of $\Sigma$.

Numerical simulations of eq.(\ref{e:gl4}) yield a surprisingly rich behavior for
small domain structures which consist only of two domains. This is discussed 
in detail in \cite{RaRi95}. Here our
interest is in the behavior of long systems containing many domains. In particular
we are interested in the interaction between adjacent domain walls.

As mentioned above, adjacent domain
walls can lock into each other due to their oscillatory interaction in space which
is related to the oscillatory behavior of the local wave number 
\cite{PeKr86,BoZi88,BoKa90,RaRi95,RaRi95a}. The solutions of (\ref{e:gl4})
are qualitatively similar to the structure shown in fig.\ref{f:typsol}. The oscillations lead to a 
discretization of the allowed domain widths, such that the domains can
be characterized by the number of oscillations contained in them. The dependence of
the range of existence of a domain on the number of oscillations has been 
studied in \cite{RaRi95} for domains ranging from 1 to 9 extrema in the wave number.
It was found that among these domains those with 5 extrema exist 
over the largest range of $D$. 

\begin{figure}
\begin{picture}(420,660)(0,0)
\put(-50,-45) {\includegraphics{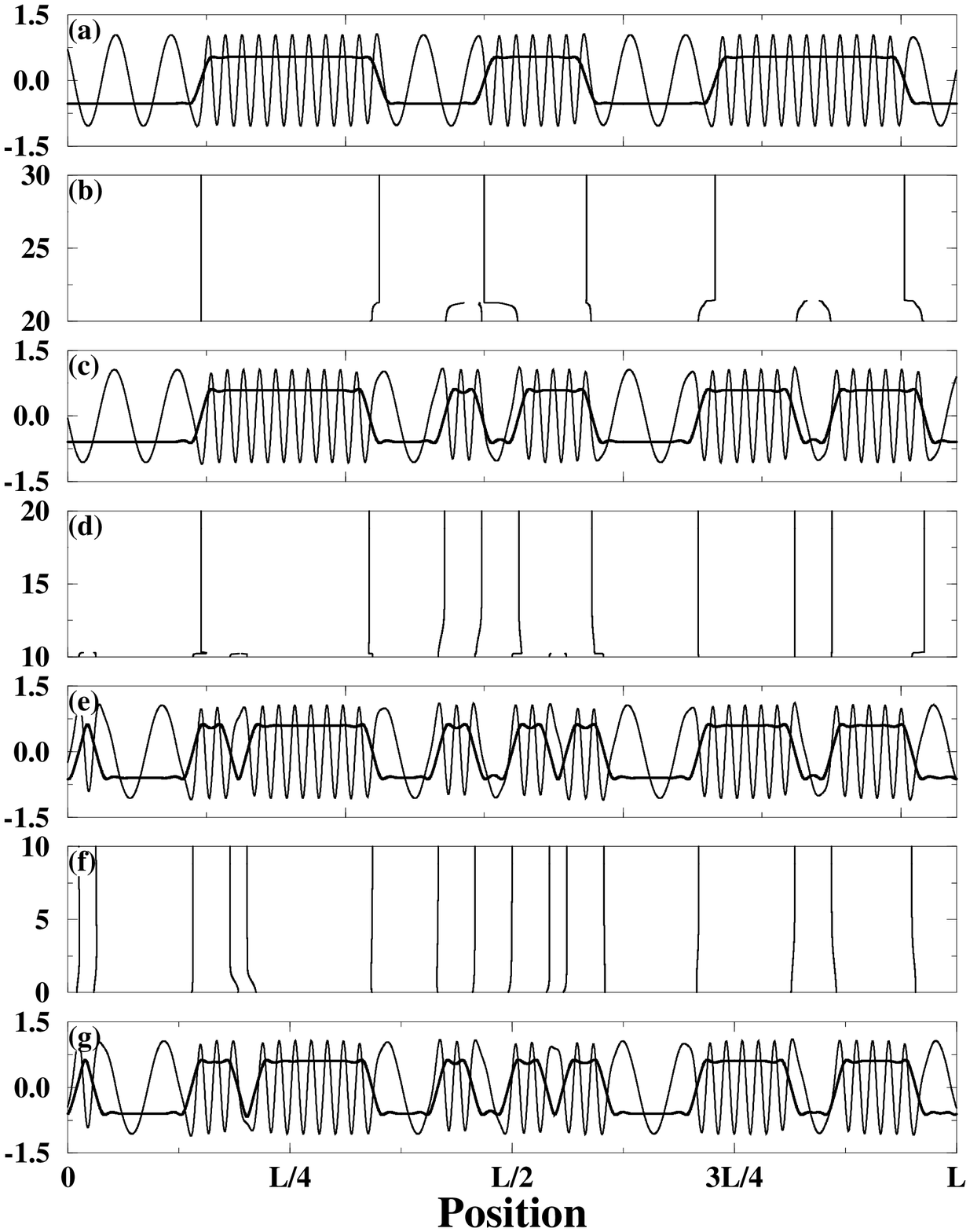}}
\end{picture}
\caption{Positions of domain walls as they evolve, and the
domain structures observed at various times.  Parts (a), (c), (e), and (g)
show the wave number (thick line) and $Re({\cal A} e^{iq_0x})$ (thin line)
while (b),(d), and
(f) plot the positions of zeros of the wave number.  For all figures,
$\Sigma=1.0$ and for (a-c) $D=-0.72$, (d,e)
$D=-0.7$ and (f,g) $D=-0.58$.  The times indicated in figures (b), (d),
and (f) are time/$10,000$.}
\protect{\label{f:dompic}}
\end{figure}

Fig.\ref{f:dompic}(g)-(a) shows the typical evolution of a large array of domains 
for periodic boundary conditions when
the coefficient $D$ is increased from negative values. An initial state
(fig.\ref{f:dompic}g) consisting of an arbitrary sequence of domains of 
different sizes is chosen.
Fig.\ref{f:dompic}f shows the temporal evolution of the zero crossings of the 
wave number at fixed $D=0.72$; clearly, domains that 
do not have the appropriate width become either wider or narrower in order
to lock the domain walls at an appropriate distance. The resulting state 
(fig.\ref{f:dompic}e) is stable.
In analogy to the analysis of spatial chaos in a 
Ginzburg-Landau equation with real amplitude \cite{CoEl87}, it is 
expected that in very long states of the type shown in fig.\ref{f:dompic}d
the sequences of domain lengths can be chaotic \cite{BoZi88,BoKa90}.

If $D$ is changed from $D=-0.72$ to $D=-0.7$ states which have only
one extremum between adjacent walls cease to exist and the adjacent domain walls
annihilate each other. Such a situation is shown in the sequence  
fig.\ref{f:dompic}(e)-(c). The single-extremum states disappear
 very soon after the control parameter is changed. The remaining states 
then reorder themselves to retain their desired spacing. A similar process is observed in the
sequence fig.\ref{f:dompic}(c)-(a) when the control parameter is changed
to $D=-0.58$, a value at which the three-extrema states no longer exist. Again, the domain
walls
annihilate each other and a new stable state consisting only of very wide domains
is reached. In the numerical simulations such states with more than 9 extrema were not
found to be stable. On an exponentially long time scale the state 
depicted in fig.\ref{f:dompic}a
is therefore expected to coarsen to a state with only one domain with large and one domain
with small wave number. 

\section{Conclusion} 
\label{s:conc}
In this paper we have investigated the stability of parametrically
driven waves in systems in which the dispersion relation for unforced waves is
non-monotonic. We have studied coupled Ginzburg-Landau equations which are
valid for small amplitudes and in the vicinity of the inflection point 
in the dispersion relation. As expected, the neutral curves for the excitation of
the waves exhibit multiple minima in this regime, and consequently the band of 
stable wave numbers is split up into three sub-bands. Numerical simulations show that
under these conditions patterns consisting of an array of domains with different
wave numbers can be stable. Our results suggest that such domain structures should
be readily accessible experimentally. Particularly suitable appear experiments on 
surface waves in ferro fluids in the presence of a static and a time-periodic magnetic 
field. There the non-monotonic dispersion relation is a precursor of the Rosensweig
instability \cite{Ro85a,Ro87}. 
Analogous experiments with conducting or dielectric fluids
in the presence of the corresponding electric fields are expected to give similar results
\cite{TaMc65,HeMi90,El94}.
 
It has been pointed out previously \cite{Ri90a,Ri90} that domain structures are also 
likely to occur if a parametric forcing is applied to waves which are 
Benjamin-Feir unstable \cite{BeFe67} in the absence of forcing. This can be seen from
(\ref{e:D}). In the limit of large amplitudes the diffusion coefficient becomes negative
in the band center if $c_r d_r+c_i d_i >0$ which is the condition for Benjamin-Feir 
instability of the unforced waves \cite{StPr78}. This implies that either the parametrically
forced waves become unstable at all wave numbers or the stable band has split into
subbands. In the latter case domain structures should arise.
 
An interesting open question concerns the stability of domain structures in two dimensions. 
If the wave numbers in the different domains differ only in their orientation
but not in their magnitude one obtains zig-zag structures. In isotropic systems they arise
generically due to the annular shape of the range of stable wave numbers. 
In axially anisotropic systems they appear when the instability of the structureless state
 is to rolls or waves oblique to the preferred direction. The stability of zig-zag
structures has been
studied in some detail \cite{RaRi95,MaNe90a}, in particular in systems with axial anisotropy 
\cite{PeKr86,BoKa90}. 

The neutral stability surfaces for parametrically driven waves in two-dimensional 
anisotropic systems have been studied in some
detail in \cite{RiSi94}. There interesting situations with multiple minima at
different orientation as well as different magnitude of the wave number have been
found suggesting the possibility of 
 a patchwork of two-dimensional domains in which the wave numbers differ not only in
their orientation but also in their
magnitude. Can such structures be stable? Some impression can be gained from a simulation
presented in \cite{CoEm92}. Fig.3(b) of \cite{CoEm92} shows the result of
a two-dimensional simulation
of the coupled Ginzburg-Landau equations for parametrically driven waves (with 
normal dispersion) in an anisotropic system. The pattern shown is clearly
characterized by domains or patches with large and small wavelengths. 
Although the authors attribute the result to a Benjamin-Feir 
instability of the waves, i.e. a situation in which  waves of {\it all} wave numbers
 become unstable, comparison with (\ref{e:D})  shows that for the parameters 
chosen in that simulation the stable band is split into two sub-bands with the 
critical wave number $q=0$ lying in the unstable region between. Only the stability of the
latter wave number had been studied in \cite{CoEm92}. 
A better understanding of such complex two-dimensional domain structures is clearly desirable.

Finally, an extension to traveling waves is also of interest.
So far, domain structures of traveling waves have been investigated only within  a
suitable phase equation \cite{BrDe92}. The possibility of a locking of the domain
walls has, however, not been adressed. 

{\it Acknowledgements.} HR gratefully acknowledges stimulating discussions with
I. Rehberg and V. Steinberg. This work 
was supported by DOE through grant (DE-FG02-92ER14303) and by
an equipment grant from NSF (DMS-9304397). 
DR is supported by the U.S. Department of Energy, contract No.
DE-FG05-95ER14566, and also in part by the Supercomputer Computations
Research Institute which is partially funded by the U.S. Department
of Energy, contract No. DE-FC05-85ER25000.

\bibliography{/home2/hermann/.index/journal}

\begin{thebibliography}{10}

\bibitem{CrHo93}
M. Cross and P. Hohenberg, Rev. Mod. Phys. {\bf 65},  851  (1993).

\bibitem{PoMa79}
Y. Pomeau and P. Manneville, J. Phys. Lett. (Paris) {\bf 23},  L609  (1979).

\bibitem{Ec65}
W. Eckhaus, {\em Studies in nonlinear stability theory} (Springer, New York,
  1965).

\bibitem{KrZi85}
L. Kramer and W. Zimmermann, Physica D {\bf 16},  221  (1985).

\bibitem{BrDe89}
H. Brand and R. Deissler, Phys.~Rev.~Lett. {\bf 63},  508  (1989).

\bibitem{Ri90a}
H. Riecke, Europhys.~Lett. {\bf 11},  213  (1990).

\bibitem{DeLe90}
R. Deissler, Y. Lee, and H. Brand, Phys.~Rev. A {\bf 42},  2101  (1990).

\bibitem{BrDe90}
H. Brand and R. Deissler, Phys.~Rev. A {\bf 41},  5478  (1990).

\bibitem{Ri90}
H. Riecke,  in {\em Nonlinear Evolution of Spatio-Temporal Structures in
  Dissipative Continuous Systems}, edited by F. Busse and L. Kramer (Plenum
  Press, New York, 1990), pp.\ 437--444.

\bibitem{HeVi92}
J. Hegseth, J. Vince, M. Dubois, and P. Berg\'e, Europhys. Lett. {\bf 17},  413
   (1992).

\bibitem{RaRi95}
D. Raitt and H. Riecke,  in {\em Spatiotemporal Patterns in Nonequilibrium
  Complex Systems}, Santa Fe Institute, edited by P. Cladis and P.
  Palffy-Muhoray (Addison-Wesley, Reading, 1995), pp.\ 255--264.

\bibitem{RaRi95a}
D. Raitt and H. Riecke, Physica D {\bf 82},  79  (1995).

\bibitem{CoEl87}
P. Coullet, C. Elphick, and D. Repaux, Phys. Rev. Lett. {\bf 58},  431  (1987).

\bibitem{PeKr86}
W. Pesch and L. Kramer, Z. Phys. B {\bf 63},  121  (1986).

\bibitem{BoZi88}
E. Bodenschatz, W. Zimmermann, and L. Kramer, J. Phys. (Paris) {\bf 49},  1875
  (1988).

\bibitem{BoKa90}
E. Bodenschatz {\it et~al.},  in {\em The Geometry of Non-Equilibrium}, edited
  by P. Coullet and P. Huerre (Plenum, New York, 1990), p.\ 111.

\bibitem{BuFi93}
F. Busse and F. Finocchi, Phys. Earth Planetary Interiors {\bf 80},  13
  (1993).

\bibitem{TaEd90}
R. Tagg, W. Edwards, and H. Swinney, Phys. Rev. A {\bf 42},  831  (1990).

\bibitem{TeAn96}
S. Tennakoon, C. Andereck, J. Hegseth, and H. Riecke, in preparation  .

\bibitem{MaGr96}
T. Mahr, A. Groisman, and I. Rehberg, J. Magn. Magn. Materials  (submitted).

\bibitem{Ro85a}
R. Rosensweig, {\em Ferrohydrodynamics} (Cambridge University Press, Cambridge,
  1985).

\bibitem{Ro87}
R. Rosensweig, Ann. Rev. Fluid Mech. {\bf 19},  437  (1987).

\bibitem{TaMc65}
G. Taylor and A. McEwan, J. Fluid Mech. {\bf 22},  1  (1965).

\bibitem{HeMi90}
J. He, N. Miskovsky, P. Cutler, and M. Chung, J. Appl. Phys. {\bf 68},  1475
  (1990).

\bibitem{SuCh93}
G. Desurgy, J. Chabrerie, O. Denoux, and J. Wesfreid, J. Phys. II {\bf 3},
  1201  (1993).

\bibitem{El94}
A. Elhefnawy, Physica A {\bf 297},  561  (1994).

\bibitem{RiCr88}
H. Riecke, J. Crawford, and E. Knobloch, Phys.~Rev.~Lett. {\bf 61},  1942
  (1988).

\bibitem{Wa88}
D. Walgraef, Europhys. Lett. {\bf 7},  485  (1988).

\bibitem{RiSi94}
H. Riecke, M. Silber, and L. Kramer, Phys. Rev. E {\bf 49},  4100  (1994).

\bibitem{Pr91}
M. Proctor, Phys. Fluids A {\bf 3},  299  (1991).

\bibitem{BeFe67}
T. Benjamin and J. Feir, J. Fluid Mech. {\bf 27},  417  (1967).

\bibitem{StPr78}
J. Stuart and R. DiPrima, Proc. R. Soc. Lond. A {\bf 362},  27  (1978).

\bibitem{MaNe90a}
B. Malomed, A. Nepomnyashchy, and M. Tribelsky, Phys.~Rev. A {\bf 42},  7244
  (1990).

\bibitem{CoEm92}
P. Coullet, K. Emilsson, and D. Walgraef, Physica D {\bf 61},  132  (1992).

\bibitem{BrDe92}
H. Brand and R. Deissler, Phys.~Rev. A {\bf 46},  888  (1992).

\end{thebibliography}


\end{document}